\documentclass[twocolumn,showpacs,preprintnumbers,nofootinbib,superscriptaddress]{revtex4-1}
\usepackage{graphicx,amsmath,amssymb,xcolor,hyperref}
\usepackage[utf8]{inputenc}
\usepackage{ulem}
\usepackage{comment}

\def\dslash{\rlap{\hspace{0.03cm}/}{\partial}}

\def\eDMEFT{e\textsc{DMeft}}

\begin{document}

\author{Giorgio Arcadi}
\email{giorgio.arcadi@uniroma3.it}
\affiliation{Dipartimento di Matematica e Fisica\\ Universit\`a di Roma 3, Via della Vasca Navale 84, 00146, Roma, Italy}
\affiliation{INFN  Sezione  Roma  Tre}
\author{Andreas Bally}
\email{andreas.bally@mpi-hd.mpg.de}
\author{Florian Goertz}
\email{florian.goertz@mpi-hd.mpg.de}
\author{Karla Tame-Narvaez}
\email{karla.tame-narvaez@mpi-hd.mpg.de}
\author{Valentin Tenorth}
\email{valentin.tenorth@mpi-hd.mpg.de}
\affiliation{Max-Planck-Institut f{\"u}r Kernphysik\\ Saupfercheckweg 1, 69117 Heidelberg, Germany}
\author{Stefan Vogl}
\email{stefan.vogl@mpp.mpg.de}
\affiliation{Max-Planck-Institut f{\"u}r Physik \\  F\"ohringer Ring 6, 80805 M\"unchen, Germany}

\title{
EFT Interpretation of XENON1T Electron Recoil Excess:\\Neutrinos and Dark Matter}

\begin{abstract} 
We scrutinize the XENON1T electron recoil excess in the scalar-singlet-extended dark matter effective field theory. We confront it with various astrophysical and laboratory constraints both in a general setup and in the more specific, recently proposed, variant with leptophilic $Z_2$-odd mediators. The latter also provide mass to the light leptons via suppressed $Z_2$ breaking, a structure that is well fitting with the nature of the observed excess and the discrete symmetry leads to non-standard dark-matter interactions.
We find that the excess can be explained by neutrino--electron interactions, linked with the neutrino and electron masses, while dark-matter--electron scattering does not lead to statistically significant improvement. We analyze the parameter space preferred by the anomaly and find severe constraints that can only be avoided in certain corners of parameter space.  
Potentially problematic bounds on electron couplings from Big-Bang Nucleosynthesis can be circumvented via a late phase transition in the new scalar sector.

\end{abstract}

\maketitle

\section{Introduction and Setup}
\label{sec:intro}

Recently, the XENON1T collaboration reported new results from an analysis of low-energy electronic recoil data~\cite{Aprile:2020tmw}. In the energy range between $1\--5$ keV the collaboration observes an excess of events that could point towards new physics. An interpretation of the data in terms of solar axions, or a neutrino magnetic moment finds substantial statistical improvements over the background model with significances above $3 \sigma$. A more mundane explanation is also possible. With the current understanding of the experiment, a contamination with tritium, which could contribute to the excess via its beta decays, cannot be excluded and it is clearly premature to celebrate the discovery of physics beyond the Standard Model (SM). Nevertheless, the excess has attracted attention and it is of great interest to scrutinize alternative explanations and identify independent experimental probes that could confirm or refute them. As the excess is observed in a handful of bins above the threshold only theories that predict a highly localized energy deposit or an IR-dominated recoil energy spectrum can account for the observation.  The proposed solutions include but are not limited to new interactions of neutrinos ~\cite{Boehm:2020ltd,AristizabalSierra:2020edu,Khan:2020vaf,Ge:2020jfn,Chao:2020yro,Gao:2020wfr}, the absorption of keV-scale dark matter (DM) \cite{Takahashi:2020bpq,Alonso-Alvarez:2020cdv,Bloch:2020uzh,Okada:2020evk,He:2020wjs}, scattering induced by a new $U(1)$ \cite{Lindner:2020kko,Bally:2020yid}, semi-relativistic or boosted DM \cite{Kannike:2020agf,Fornal:2020npv,Su:2020zny,Cao:2020bwd,Alhazmi:2020fju}, axions \cite{DiLuzio:2020jjp,Gao:2020wer}, inelastic DM scattering~\cite{Harigaya:2020ckz,Bell:2020bes,Lee:2020wmh,Bramante:2020zos,Baek:2020owl}, large neutrino magnetic moment \cite{Babu:2020ivd,Shoemaker:2020kji,Miranda:2020kwy} or more exotic explanations \cite{McKeen:2020vpf,Zu:2020idx,Jho:2020sku,Ko:2020gdg,Cacciapaglia:2020kbf}.
It is also an interesting possibility that the energy deposit is in the form of photons since the XENON1T analysis does not differentiate between photons and electron recoils \cite{Paz:2020pbc}. However, nuclear excitations, which naturally lead to such a signal, are not  a viable solution since the energy scale does not match the observations \cite{Arcadi:2019hrw,McCabe:2015eia}.

In this article, we characterize the XENON1T excess in an effective field theory (EFT) description of a dark sector, recently proposed in \cite{Goertz:2019wtt}, 
which naturally includes the appropriate ingredients for its explanation, namely modified neutrino interactions with electrons via a potentially light new scalar sector, coupling most prominently to the  light fermion generations. The overarching framework is provided by the scalar-mediator extended DM EFT (\eDMEFT) \cite{Alanne:2017oqj,Alanne:2020xcb}, which corresponds to a systematic EFT incarnation of simplified DM models,
with a modest number of new free parameters. In the variant employed here, 
a spontaneously broken $Z_2$ symmetry is included, that on the one hand leads to interesting DM signatures \cite{Goertz:2019wtt}, and on the other allows to address the smallness of first-generation fermion masses 
via small symmetry-breaking effects, as detailed below. This automatically induces non-trivial couplings of the new scalar with these fermions and thereby allows to relate the XENON1T excess with the observed electron and neutrino masses.
While the setup features in general a suppression of the DM direct-detection (DD) cross section due to the $Z_2$ symmetry, the leptophilic variant we will consider below is completely unconstrained from DD limits via nucleon interactions and thus invites searches employing electronic recoil.

The paper is organized as follows. In the remainder of this section we detail the setup sketched above. In Section \ref{sec:XE} we provide our fit to the XENON1T electron recoil excess, first assuming neutrino-electron scattering as its origin and then exploring in addition the option of DM-scattering. Here, we will also examine whether the correct relic abundance can be achieved consistently with the tentative XENON1T signal. Subsequently, in Section \ref{sec:Cons}, we will confront the explanation with stringent limits on new electron and neutrino interactions from terrestrial and astrophysical constraints, considering also the fully agnostic case of freely modified couplings in a simple EFT and characterizing viable regions.  In Section \ref{sec:PT}, we will show how our extended scalar sector allows to circumvent stringent constraints from Big-Bang Nucleosynthesis (BBN) via a late phase transition before presenting our conclusions in Section \ref{sec:Conc}.

\subsection{General Setup}

We consider the leptophilic variant of the \eDMEFT\, recently put forward in \cite{Goertz:2019wtt},
which corresponds to the SM field content augmented with a fermionic DM singlet $\chi$ and a real, CP even scalar mediator ${\cal S}$,
with the assumption that ${\cal S}$ and the right-handed first lepton generation are odd under a $Z_2$ parity. In this model, the first-generation leptons obtain masses from the vacuum expectation value (vev) of ${\cal S}$. 

We will here extend the scalar sector by assuming two different $Z_2$ symmetries, one shared by the neutrinos, the other by the electron, that are then broken by vevs of two distinct scalars ${\cal S}_{\nu,e}$ and allow to simultaneously address tiny neutrino masses and small charged-lepton masses. As we will show in Section \ref{sec:PT}, this richer scalar sector can also lead to a delayed coupling of the electrons to the lighter mediator in the thermal evolution of the universe, which will make possible to address the recoil excess with moderately sizable electron couplings, while avoiding bounds from physics of the early universe.

The corresponding Lagrangian reads~\cite{Goertz:2019wtt}
\begin{align}   \label{eq:LEFT}
    \mathcal{L}_{\rm eff}^{{\cal S} \chi} &= {\cal L}_{\rm SM^\prime} + \frac{1}{2} \left( \partial_\mu {\cal S}_\ell \partial^\mu {\cal S}_\ell - \mu_\ell^2 {\cal S}_\ell ^2 \right) + \bar \chi i \dslash\chi- m_\chi \bar \chi \chi \nonumber \\
    &- \frac{\lambda_\ell}{4} {\cal{S}}_\ell^4 - \lambda_{\nu e}\, {\cal{S}}_\nu^2 {\cal{S}}_e^2\ - \lambda_{H{\cal S}_\ell} |H|^2 {\cal S}_\ell^2\\
    &- \frac{1}{\Lambda} \Big[(y_\nu^{\cal S})_{ij}\, \bar{L}_L^i \tilde H \nu^j_R\, {\cal S}_\nu + (y_e^{\cal S})_i\, \bar{L}_L^i H e_R\, {\cal S}_e +  \mathrm{h.c.}\Big]\nonumber\\[1mm]
    &- \frac{y_\chi^{\ell} {\cal S}_\ell^2 + y^H_\chi |H|^2}{\Lambda}\ \bar{\chi}_L \chi_R + \mathrm{ h.c.}\,,\nonumber
\end{align}
where a summation over $\ell=\nu,e$ is understood and $L_L$ are the left-handed $SU(2)_L$ lepton doublets, $e_R, \nu^e_R$ are the right-handed electron and right-handed neutrinos, while $H$ is the Higgs doublet.
The latter develops a vev, $|\langle H \rangle| \equiv v/\sqrt 2 \simeq 174$\,GeV, triggering electroweak symmetry breaking (EWSB). In unitary gauge, the Higgs field is expanded around the vev as $H \simeq 1/\sqrt2 (0, v + h)^T$, where $h$ is the physical Higgs boson with mass $m_h \approx 125$\,GeV. Finally, ${\cal L}_{\rm SM^\prime}$ denotes the SM Lagrangian without the Yukawa couplings of the electron (and the neutrinos), see Eq.~\eqref{eq:Lm} below.
Importantly, also the mediators develop small vevs $|\langle {\cal S}_\ell \rangle| \equiv  v_\ell \ll v $, which break the $Z_2^{\ell}$ symmetries, $\ell=\nu, e$, carried by all the right-handed neutrinos and the right-handed electron, respectively, and thereby generate masses for the light leptons.
Even though we work in the EFT framework in order to focus on the broad characteristics of the XENON1T excess, we present two explicit UV completions that can lead to the crucial terms in the Lagrangian~\eqref{eq:LEFT} in Appendix \ref{sec:UV}.

The mixing with the Higgs via the $|H|^2 {\cal S}_\ell^2$ operators has to be small and this effect will not be considered in the following. Note that the conventional DM interaction ${\cal S}_\ell\, \bar{\chi}\chi$ is still generated with coefficient $\sim 2 y_\chi^\ell v_\ell/\Lambda$ which will remain relevant for our analysis.
Finally, we take the coefficient of the operator $|H|^2 \bar{\chi}\chi$ to be negligibly small, such as to evade direct detection constraints and limits from invisible Higgs decays \cite{Fedderke:2014wda}.

\subsection{Fermion Masses, Scalar Mixing, and Free Parameters}

In the following we will study the fermion and scalar mass spectrum of the setup and summarize its relevant free parameters.
The fermion mass terms after electroweak and $Z_2^\ell$ breaking read
\begin{equation}
	{\cal L} \supset - \!\sum_{\ell=e,\nu}\! \bar \ell_L \frac{v}{\sqrt 2} \left( Y_\ell^H + \frac{v_\ell}{\Lambda} Y_\ell^{\cal S} \right) \ell_R \equiv - \!\sum_{\ell=e,\nu}\! \bar \ell_L M^\ell \ell_R \,,
	\label{eq:Lm}
\end{equation}
where $\ell_{L,R}=e_{L,R},\nu_{L,R}$ are three-vectors in flavor space and the Yukawa matrices
\begin{equation} 
\label{eq:Yukawas}
\begin{split}
	Y_e^H = \begin{pmatrix}
	0 & y_{12}^e & y_{13}^e \\
	0 & y_{22}^e & y_{23}^e \\
	0 & y_{32}^e & y_{33}^e
	\end{pmatrix}\,,\quad	
	\quad Y_e^{\cal S} =  \begin{pmatrix}
	({y_e^{\cal S}})_1 & 0 & 0 \\
	({y_e^{\cal S}})_2 & 0 & 0\\
	({y_e^{\cal S}})_3 & 0 & 0
	\end{pmatrix}\,,
	\\
	Y_\nu^H = \text{\bf 0}\,,\hspace{1cm} 	
	Y_\nu^{\cal S} =  \begin{pmatrix}
	({y_\nu^{\cal S}})_{11} & ({y_\nu^{\cal S}})_{12} & ({y_\nu^{\cal S}})_{31} \\
	({y_\nu^{\cal S}})_{21} & ({y_\nu^{\cal S}})_{22} & ({y_\nu^{\cal S}})_{32}\\
	({y_\nu^{\cal S}})_{31} & ({y_\nu^{\cal S}})_{32} & ({y_\nu^{\cal S}})_{33}
	\end{pmatrix}\,,
\end{split}
\end{equation}	
reflect the $Z_2^\ell$ assignments. Without breaking of the latter symmetry via $v_{ \ell}>0$, the electron and the neutrinos would remain massless, corresponding to vanishing eigenvalues of $Y_\ell^H$. On the other hand, a small breaking of $v_\nu \sim {\cal O}({\rm eV})$ and $v_e \sim {\cal O}({\rm MeV})$ is sufficient to generate $m_\nu \sim 0.1$\,eV and $m_e \sim 0.5$\,MeV with natural $({y_\ell^{\cal S}}) \lesssim {\cal O}(1)$ and $\Lambda \gtrsim 1\,$TeV. We note that to explain the XENONT1T excess in light of various constraints, it will later be necessary to somewhat deviate from these natural scales, remaining with a partly explanation of light-fermion masses.

We now perform a rotation to the mass basis
\begin{equation}
\begin{split}
	M^\nu &= U_L^\nu\, M_{\rm diag}^\nu U_R^{\nu\, \dagger},
	\ \, M_{\rm diag}^\nu\!={\rm diag}(m_{\nu^1},m_{\nu^2},m_{\nu^3})\,,\\
	M^e &= U_L^e\, M_{\rm diag}^e U_R^{e\, \dagger},
	\ \, M_{\rm diag}^e\!={\rm diag}(m_e,m_\mu,m_\tau)\,,
\end{split}
\end{equation}
with $U_L^e = U_L^\nu\, V_{\rm PMNS}$, to obtain the couplings of the physical leptons to the Higgs boson and the scalar mediators \cite{Goertz:2019wtt}
\begin{equation}
	{\cal L} \supset - \sum_{\ell=e,\nu} \bar \ell_L \left(  \frac{\hat{Y}_\ell^H + v_\ell/\Lambda\, \hat{Y}_\ell^{\cal S}}{\sqrt 2}\, h + \frac{v\, \hat{Y}_\ell^{\cal S}}{\sqrt 2 \Lambda}\, {\cal S}_\ell \right) \ell_R\,,
\end{equation}
where $\hat{Y}_\ell^s = U_L^{\ell\, \dagger}  Y_\ell^s U_R^\ell,\, s=H,{\cal S};\,\ell=\nu,e$, and (with some abuse of notation) we denote the mass eigenstates by the same spinors $\ell=e,\nu$.
The Yukawa matrices in the mass basis can be expressed as 
\begin{equation} 
\begin{split}
	\hat Y_\ell^{\cal S} =& \frac{\sqrt 2 \Lambda}{v v_\ell} \, M_{\rm diag}^\ell \, U_R^{\ell\, \dagger} C_\ell^S\, U_R^\ell\,, \\[2mm]
	\hat Y_\ell^H =& \frac{\sqrt 2}{v} \,M_{\rm diag}^\ell\, U_R^{\ell\, \dagger} C_\ell^H \,U_R^\ell \,,
\end{split}
\end{equation}
where $C_e^S = {\rm diag}(1,0,0)\,, C_\nu^S ={\rm diag}(1,1,1) \,,  C_e^H = {\rm diag}(0,1,1)\,, C_\nu^H = \text{\bf 0}$, and the unitary rotations of the left-handed lepton fields drop out since they share the same $Z_2^{\ell}$ charges and their couplings (with a fixed right-handed lepton) are thus aligned with the corresponding mass terms. While this is not true for the right handed leptons, which could introduce flavor-changing neutral currents (FCNCs), here we just chose the Yukawa matrices starting from $M_{\rm diag}^\ell$, such that $U_R^e = {\bf 1}$, avoiding FCNCs \cite{Goertz:2019wtt}. We thus finally arrive at 
\begin{equation} 
\label{eq:YukFin2}
\begin{split}
    \hat Y_\nu^{\cal S} =& \,\frac{\sqrt 2 \Lambda}{v v_\nu}  
	\,{\rm diag}(m_{\nu^1},m_{\nu^2},m_{\nu^3})\,,\\
	\hat Y_e^{\cal S} =& \, \frac{\sqrt 2 \Lambda}{v v_e} \,{\rm diag}(m_e,0,0)\,,\\
	\hat Y_\nu^H =&\ \text{\bf 0}\,,\\
	\hat Y_e^H =&\, \frac{\sqrt 2}{v}\, {\rm diag}(0,m_\mu,m_\tau)\,.
\end{split}
\end{equation}

In consequence, the $\mu$ and $\tau$ interact with the Higgs boson as in the SM while the electron couples instead only to ${\cal S}_e$, with strength determined by the free parameter $v_e$, which can be traded for ${\rm y}_e^{\cal S}/\Lambda \equiv (\hat Y_e^{\cal S})_{11}/\Lambda$. Similarly, $v_\nu$ can be traded e.g. for ${\rm y}_1^{\cal S}/\Lambda$, where ${\rm y}_i^{\cal S}/\Lambda \equiv (\hat Y_\nu^{\cal S})_{ii}/\Lambda$.

In addition to fermion mixing, the scalar potential term $\sim \lambda_{\nu e}$ leads to a mixing between the scalar singlets after they obtain their vevs 
$|\langle {\cal S}_\ell \rangle| =  v_\ell$, described by an angle $\theta$ as
\begin{align}
\label{eq:rotation}
	\begin{pmatrix} s \\ S \end{pmatrix} =  
	\begin{pmatrix} \cos\theta & \sin\theta \\ -\sin\theta & \cos\theta \end{pmatrix}
	\begin{pmatrix} {\cal S}_\nu \\ {\cal S}_e \end{pmatrix}\,,
\end{align}
with
\begin{equation}
\label{eq:mixinghS}
    \tan 2\theta=\frac{4 \, \lambda_{\nu e} v_\nu v_e}{M_\nu^2 -M_e^2}\,,
\end{equation}
where $M_\ell^2 = \mu_\ell^2 + 3 \lambda_\ell v_\ell^2 + 2\lambda_{\nu e}\, v_\nu^2 v_e^2 /v_\ell^2$.
The resulting physical masses read
\begin{equation}
    m_{s/S}^2 = \frac{1}{2}(M_\nu^2 +M_e^2) \pm \frac{M_\nu^2-M_e^2}{2\,\cos 2\theta} \,.
\end{equation}
Note that, as suggested by the lightness of neutrino masses compared to the charged lepton masses, we will always assume $v_\nu \ll v_e$ and accordingly $M_\nu \ll M_e $. This leads to $m_s \approx M_\nu,\ m_S \approx  M_e $, as well as $\cos \theta \approx 1$, $\sin \theta \ll 1$.

The mixing will induce suppressed couplings between the electron and the light mediator $s$ (as well as between the neutrinos and the heavy $S$), given by
\begin{equation}
\begin{split}
    \mathcal{L}_s =& -s \frac{v}{\sqrt 2\Lambda}  \left(c_\theta {\rm y}_i^{\cal S}\, \bar{\nu}_L^i \nu^i_R + s_\theta  {\rm y}_e^{\cal S}\, \bar{e}_L e_R \right)\,,\\[2mm]
    \mathcal{L}_S =& -S \frac{v}{\sqrt 2 \Lambda} \left(c_\theta {\rm y}_e^{\cal S}\, \bar{e}_L e_R - s_\theta {\rm y}_i^{\cal S}\, \bar{\nu}_L^i \nu^i_R \right)\,,
\end{split}
\end{equation}
where ${c_\theta\equiv\cos\theta},\, {s_\theta\equiv\sin\theta}$.

Before moving on, we are now in a position to summarize the free parameters of our setup, relevant for our study, which are
\begin{itemize}
\item{the mediator masses  $m_{s,S}\approx M_{\nu,e}$}
\item{the ${\cal S}_e-$Yukawa coupling ${\rm y}_e^{\cal S}/\Lambda$}
\item{the ${\cal S}_\nu-$Yukawa coupling ${\rm y}_1^{\cal S}/\Lambda$}
\item{the mixing portal $\lambda_{\nu e}$\,,}
\end{itemize}
as well as, in the dark sector,
\begin{itemize}
\item{the DM mass $m_{\chi}$}
\item{the bi-quadratic DM portal coupling $y_{\chi}^\ell/\Lambda$}\,,
\end{itemize}
where the remaining Yukawa couplings are given by \eqref{eq:YukFin2}.

\section{Fitting the XENON1T Excess}
\label{sec:XE}

\subsection{Modified Neutrino Interactions}
\begin{figure}
    \centering
    \includegraphics[width=0.48\textwidth]{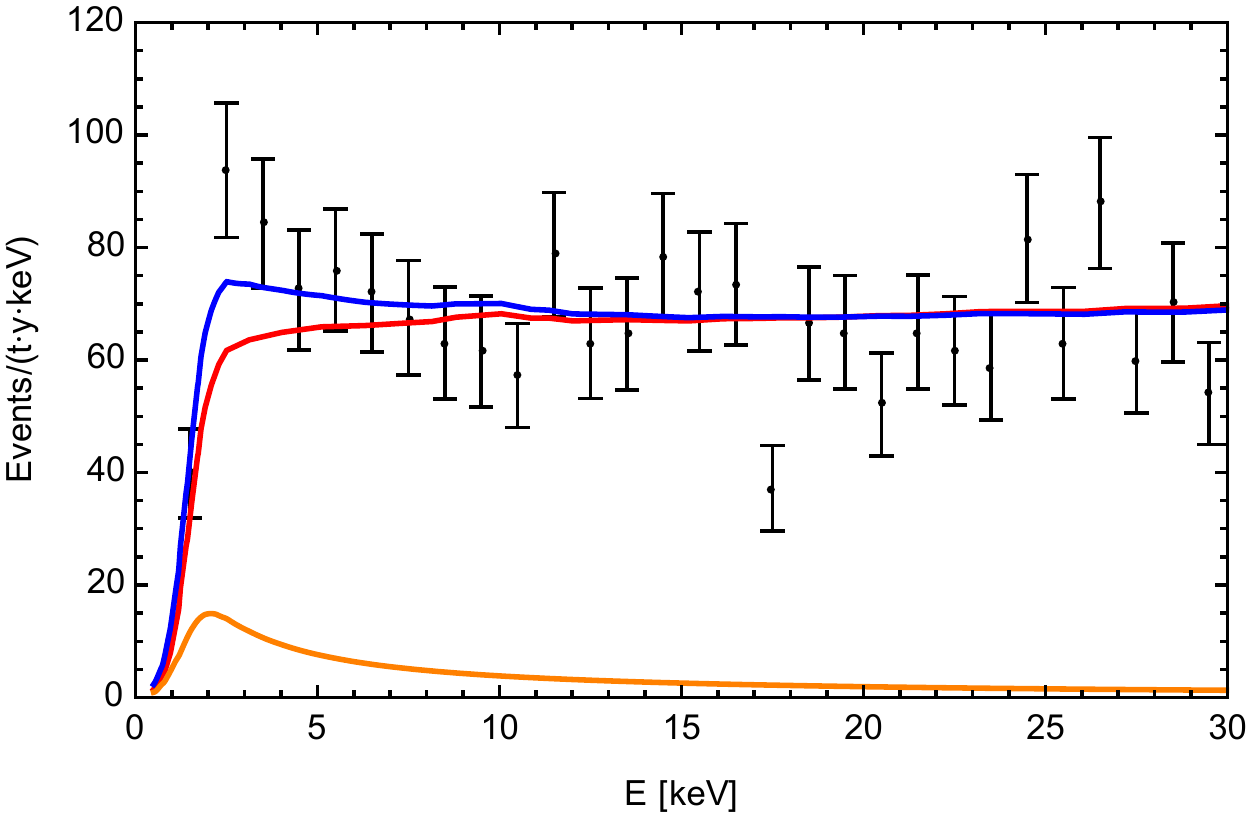}
    \caption{Comparison between an exemplary differential event rate for a scalar with $m_s=60$\,eV and $\sqrt{y^s_e y^s_\nu }=7.9 \times 10^{-7}$ and the data as reported by \cite{Aprile:2020tmw}. The full differential event rate is shown in blue while the pure signal (background) contribution is depicted in orange (red).}
    \label{fig:event_rate}
\end{figure}
We start by assuming that the $\chi$-scattering induced electron recoil is negligible. In this case, the excess can be explained by modified neutrino scattering with electrons, in our model mediated by $s$ and $ S$. As we will see observational constraints prefer $m_{s}\ll m_{\cal S}$ such that neutrino-electron scattering can to good approximation be described by $s$-exchange alone. The differential cross section for the new-physics signal reads \cite{Cerdeno:2016sfi}
\begin{equation}
    \frac{d\sigma_{\nu e}}{d E_r}= \frac{(y^s_{e} y^s_{\nu})^2}{4\pi (2 m_e E_r +m^2_s)^2}\frac{m_e^2 E_r}{E_\nu^2}\,, 
\end{equation}
where $m_e$ is the electron mass, $E_\nu$ the energy of the incoming neutrino, $E_r$ the electron recoil energy, and we denoted the couplings of the electrons and the first neutrino to the light (and heavy) mediators $s$ (and $S$) by
\begin{align}
\begin{aligned}
    y^s_e &\equiv s_\theta \frac{v}{\sqrt 2\Lambda} {\rm y}_e^{\cal S} \,,&\qquad  y^s_\nu &\equiv c_\theta \frac{v}{\sqrt 2\Lambda} {\rm y}_1^{\cal S}\,,\\
    y^S_e &\equiv c_\theta \frac{v}{\sqrt 2\Lambda} {\rm y}_e^{\cal S} \,,&\qquad y^S_\nu &\equiv -s_\theta \frac{v}{\sqrt 2\Lambda} {\rm y}_1^{\cal S}\,.
\end{aligned}
\end{align}
The true differential event rate is then given by convoluting the differential cross section and the incident neutrino flux $\phi_{\nu}$ and weighting by the number of electrons per unit mass $N_e$
\begin{equation}
    \frac{d R}{d E_r}= N_e \int\! d E_\nu\, \frac{d\sigma_{\nu e}}{d E_r}\, \frac{d\phi_{\nu}}{d E_\nu}\,.
\end{equation}
At the energies relevant for the XENON1T excess the neutrino flux is dominated by $pp$ neutrinos from the sun. We use the observed value of the $pp$-flux from \cite{Vitagliano:2019yzm} and employ the parameterization of the spectrum from \cite{Bahcall:1997eg}. Here we have assumed a universal interaction between $s$ and the different neutrino flavors such that oscillation effects do not affect the scattering rate, on which we will comment more further below.  To make connection with the observed rate experimental effects have to be included. The limited detector resolution is taken into account via a gaussian smearing function with an energy dependent resolution. As suggested in \cite{XENON:2019dti} we take the ansatz
\begin{equation}
 \sigma(E)/E=\frac{a}{\sqrt{E}}+b   
\end{equation} and assume that the resolution varies between $\approx 30\%$ at $E_r=1$ keV and $\approx 6\%$ at 30 keV. Finally, the detector efficiency reported in the experimental paper \cite{Aprile:2020tmw} is applied. We adopt the best fit background model from the experimental publication but allow the normalization to vary within the $1\sigma$ allowed range.  
In order to assess the impact of a light scalar on the electron neutrino scattering we perform a $\chi^2$ analysis of the signal and background model in light of the observed data.  

We find that a coupling of 
\begin{equation}\label{eq:fit}
    \sqrt{y^s_e\, y^s_\nu }\approx7.9 \times 10^{-7}
\end{equation}
is preferred with very little dependence on $m_s$ for masses smaller than $\approx 20$ keV.  An exemplary comparison between the signal associated with the best fit point for $m_s= 60$ eV and the data is shown in Fig.~\ref{fig:event_rate}. This choice of parameters corresponds  to  $\chi_{best}^2=38.9$ compared to $\chi^2_{bd}=47.1$ for the background-only hypothesis.

Our results are in good qualitative agreement with those in \cite{Boehm:2020ltd,Khan:2020vaf,AristizabalSierra:2020edu} which study a related set-up and we will confront them with a comprehensive set of complementary experimental constraints in Section \ref{sec:Cons}.

\subsection{DM Scattering and Relic Abundance}

\begin{figure}    \centering
    \includegraphics[width=0.48\textwidth]{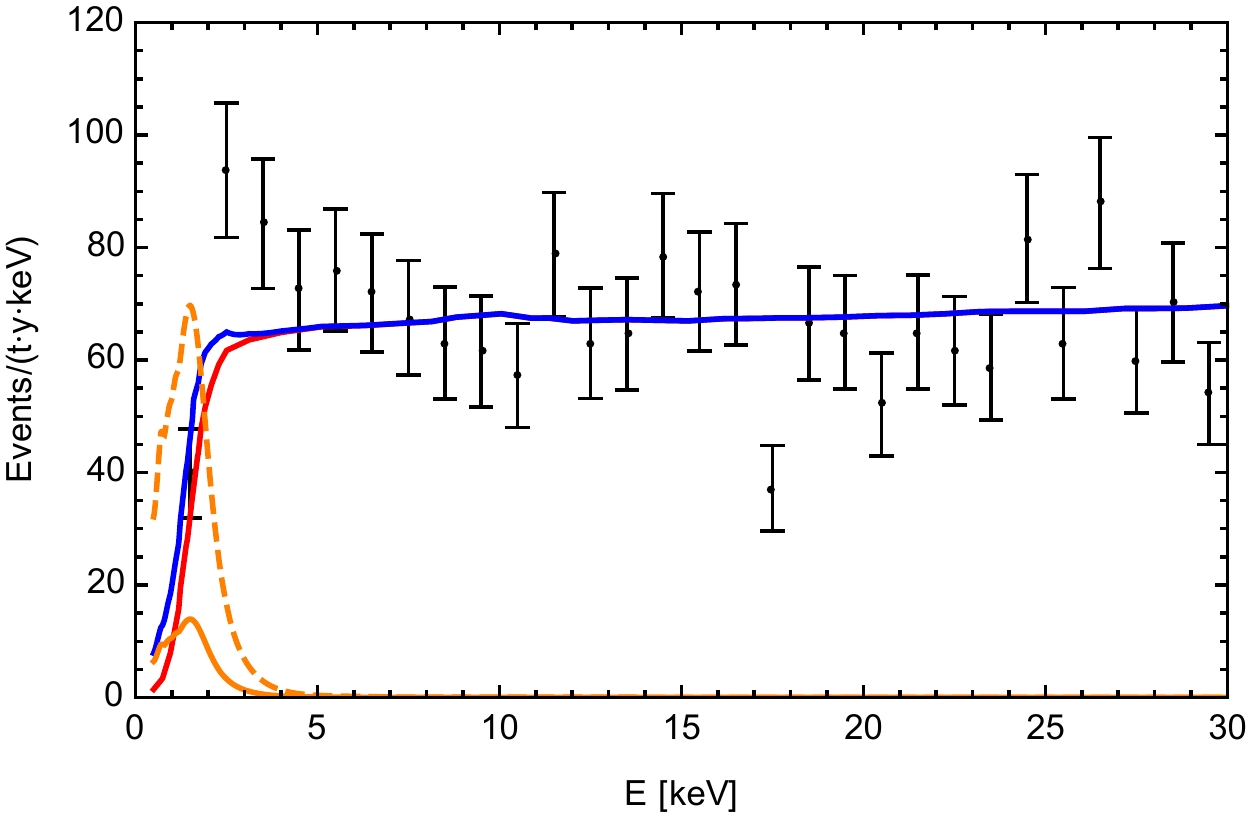}
    \caption{Comparison between the best fit differential event rate for a DM particle with $m_\chi=10$ GeV and $\sigma_{e\chi}= 1.25 \times 10^{-39}\, \mbox{cm}^2$ and the data. The style is similar to Fig.~\ref{fig:event_rate} and for better visibility we also show the signal rate enhanced by a factor of 5 as an orange dashed line.}
    \label{fig:event_rate_DM}
\end{figure}

DM can scatter on the electrons in the detector and could account for the excess. 
Beyond that, our model contains in any case a DM candidate and thus it is interesting to check
whether the correct relic abundance can be achieved simultaneously with an explanation of the XENON1T excess.
These observables are correlated with each other also in case of the neutrino explanation, via the mediator couplings to SM fermions.

A naive estimate of the maximum recoil energy possible in non-relativistic DM-electron collisions leads to
\begin{align}
    E_{r,max}= \frac{2\mu_{\chi,e}^2}{m_e}\, v^2_{max}\approx 2 \times 10^{-6} m_e \,,
\end{align}
where $\mu_{\chi,e}$ is the reduced mass of the system and $v$ the velocity of the DM. For $m_{DM}\!\gg\! m_e$ and after taking into account that the velocity is limited by the local escape velocity of our galaxy $v_{esc}\!=\mathcal{O}(10^{-3}\,c)$ this leads to an estimate of $E_{r,max} \approx 1$\,eV and thus much below the energy scale required to account for the signal. However, it is crucial to note that the electrons form part of a bound system, the xenon atom.
Therefore, the momentum of the electron is not zero but could take an arbitrary value. The typical momentum of the bound electron is expected to be $\mathcal{O}(\alpha_{em} m_e)$ which is still small but allows for a larger energy transfer in the DM-electron scattering process \cite{Roberts:2015lga}. 
The differential event rate is given by
\begin{align}
    \frac{d R}{d E_r}=\frac{n_{Xe} \rho_\chi}{m_\chi}\frac{d \langle \sigma_{\chi e}\rangle}{d E_r}
\end{align}
where $N_{Xe}$ is the number of xenon atoms per unit mass in the detector and $\rho_\chi \approx 0.3\, \mbox{GeV}/\mbox{cm}^3$ the local DM density. For the velocity averaged differential cross section 
we rely on the results of \cite{Roberts:2016xfw,Roberts:2019chv}. In the heavy mediator limit\footnote{A light mediator leads to a much stronger energy dependence of the signal and is expected to provide a worse fit of the signal than a heavy mediator.} it can be parametrized as
\begin{align}
    \frac{d \langle \sigma_{\chi e}\rangle}{d E_r} =\frac{\sigma_{\chi e}}{2 m_e} \int\! dv\, \frac{f(v)}{v}\int\! dq \,a_0^2 q K(E_r,q) \,,
\end{align}
where $\sigma_{\chi e}$ is the cross section for scattering on a free electron with a momentum transfer $a_0^{-1}=\alpha_{em} m_e$, while $f(v)$ denotes the velocity distribution of the DM at Earth. The atomic physics is encoded in the excitation factor $K$ which has originally been computed in \cite{Roberts:2016xfw}.  In order to estimate the implications of a DM signal we consider the averaged cross sections reported in~\cite{Roberts:2019chv} and perform a fit to the signal using the same assumptions about the detector as before. The best fit recoil rate we found is shown in Fig.~\ref{fig:event_rate_DM}. As can be seen the signal rises very steeply at low energies such that the peak occurs at $\approx 1.5$ keV instead of the $\approx 2.5$ keV  needed to reproduce the data. It is interesting to note that the fit shows some improvement if a small DM signal is added. The best fit corresponds to $m_\chi=10$~GeV and $\sigma_{\chi e}\approx 1.25 \times 10^{-39} \mbox{cm}^2$, which could for instance be explained by an MeV scale mediator with an $\mathcal{O}(1)$ coupling to DM and $y^S_e\approx10^{-5}$. However, the statistical improvement only amounts to marginally more than $1\sigma$. Therefore, the DM-electron-scattering hypothesis does not provide a convincing explanations of the observation and  we do not entertain this possibility  further; similar conclusions were reached for example in \cite{Bloch:2020uzh}. A better fit with DM requires a flatter recoil spectrum. This could for instance be achieved if a relativistic or semi-relativistic DM sub-population~\cite{Fornal:2020npv,Kannike:2020agf,Alhazmi:2020fju} contribute to the signal or if the interaction has additional momentum dependence \cite{Bloch:2020uzh}. 
Nevertheless, it is interesting to ask whether the observed DM relic density can be accounted for in our framework. 

Assuming production via freeze-out,  the correct relic abundance can be achieved if the thermally averaged annihilation cross-section is  $\mathcal{O}(10^{-26}\,{\mbox{cm}}^3\,{\mbox{s}}^{-1})$. In our model the main annihilation channels for the DM are into $e^+ e^-$ and $SS$ final states.\footnote{Here we neglect the corresponding contributions involving the light mediator $s$ for simplicity, which will not lead to qualitative changes.} The cross-section for the former channel can be estimated as (see, e.g.~\cite{Alanne:2020xcb}):
\begin{align}
    &\langle \sigma v \rangle_{ee} \approx \frac{1}{8\pi}\, \frac{v^2 v_e^2}{\Lambda^4}\, \frac{(y_\chi^e)^2 (y^{\cal S}_e)^2 m_\chi^2}{(m_S^2-4 m_\chi^2)^2}\ v_\chi^2 \nonumber\\
    &\approx 10^{-5} \sigma_v^0 {\left(\frac{v_e}{5\,\mbox{GeV}}\right)}^2  {\left(\frac{1\,\mbox{GeV}}{m_\chi}\right)}^2 {\left(\frac{10\,\mbox{TeV}}{\Lambda}\right)}^4 (y_\chi^e)^2 (y^{\cal S}_e)^2 \,,
\end{align}
where $v_\chi$ is the DM velocity (the cross-section is p-wave suppressed), while $\sigma_v^0=2 \times 10^{-26}\,{\mbox{cm}}^3\,{\mbox{s}}^{-1}$.
A similar estimate for the cross-section into the $SS$ final state leads to:
\begin{equation}
    \langle \sigma v \rangle (\bar\chi\chi \rightarrow SS) \approx 
    10^{-3}\, \sigma_v^0\, {\left(\frac{10\,{\rm TeV}}{\Lambda}\right)}^2 (y_\chi^e)^2.
\label{eq:xsec_XXSS}
\end{equation}
While the former cross section is way too small in case the XENON1T excess should be explained consistently, the second cross section could in principle lead to a viable scenario, however only in case $\Lambda$ is lowered to the TeV scale. Values $y_\chi^e \gtrsim 1$ would be nevertheless required.

An alternative production mechanism that is more easily realized within the setup at hand is freeze-in.
In this case  $y_\chi^\ell \ll 1$ and the DM interactions are so weak that thermal equilibrium has never been realized in the early Universe. Then the relic density can be built up from a  negligible initial value, by $SS (ss) \to \bar\chi \chi$ inverse annihilation processes and, for sufficiently light $\chi$, $S \to \bar\chi\chi$ decays.
Since it is realized via a $D=5$ operator, the annihilation process leads to a UV dominated rate. Hence the relic density is sensitive to the largest temperature and we need to specify our assumed value for the reheating temperature $T_R$. In order not to exceed the validity of our EFT we limit ourselves to $T_R$ below the new physics scale $\Lambda$.

\begin{figure}    \centering
    \includegraphics[width=.48\textwidth]{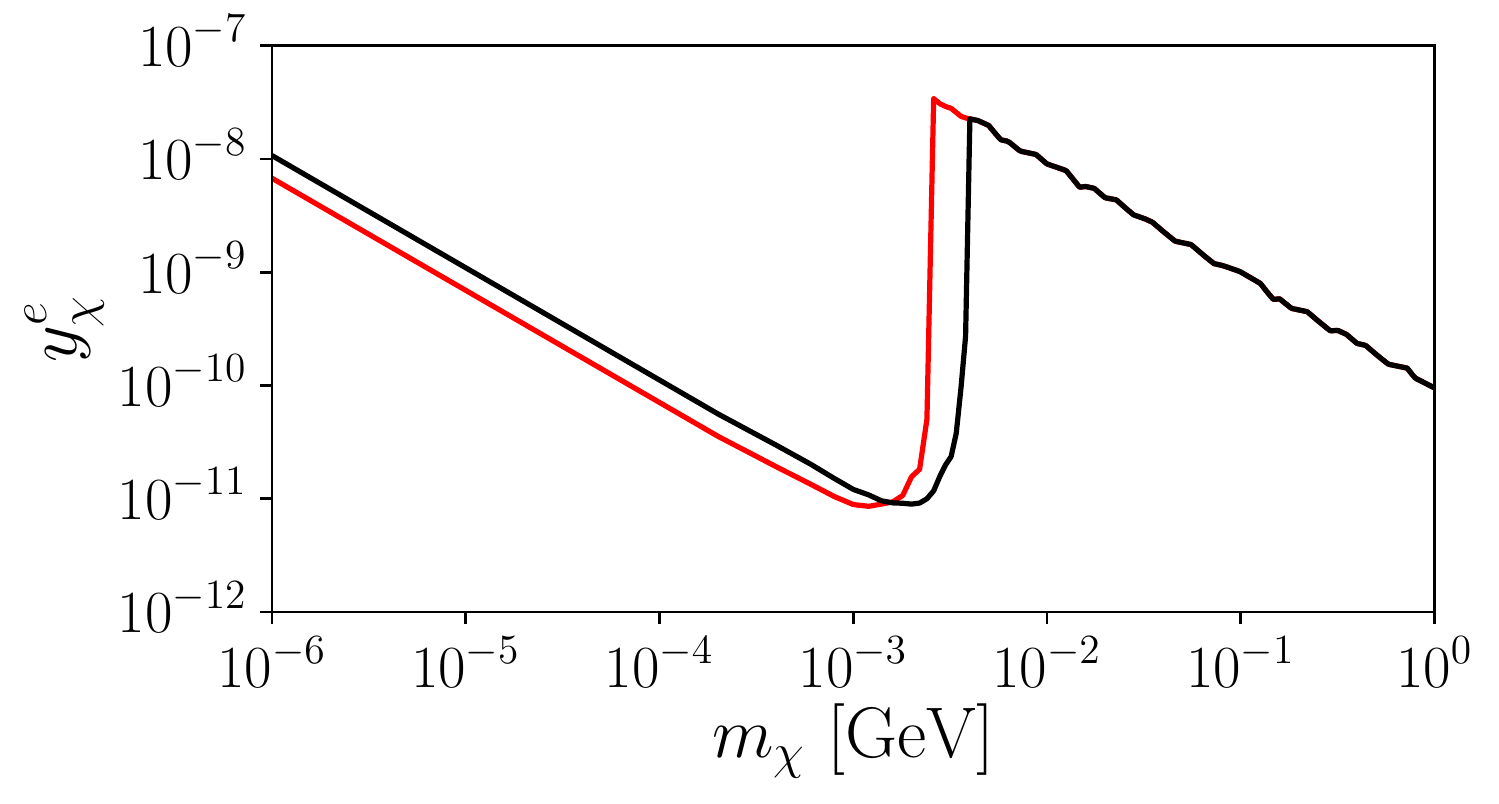}
    \caption{Isocontours of correct DM relic density assuming production through freeze-in and considering the assignations of model parameters for BM1 (red) and BM2 (black). The reheating temperature $T_R$ has been set to 100~GeV.}
    \label{fig:relic_FI}
\end{figure}

We compute the relic density with the freeze-in module of the public code micrOMEGAs 5~\cite{Belanger:2018mqt} which takes the full momentum dependence of the annihilation and decay rates into account. Fig.~\ref{fig:relic_FI} shows isocontours of $\Omega_\chi h^2 =0.12$ in the $(m_\chi,y_\chi^e)$ plane for the two  benchmark models described in the next section, assuming $y_\chi^\nu=y_\chi^e$. In our computation we have adopted $T_R=100\,\mbox{GeV}$. The DM relic density depends, besides $(m_\chi,y_\chi^e)$, on $m_S$ and $v_e$. The values of these two parameters are comparable for our benchmarks ($m_S\sim 5$\,MeV, $v_e\sim 5$\,GeV), so that the two contours in Fig.~\ref{fig:relic_FI} are rather close to each other and align in the high and low mass limit.

\section{Terrestrial and Astrophysical Constraints}
\label{sec:Cons}

\begin{figure}
    \centering
    \includegraphics[width=.48\textwidth]{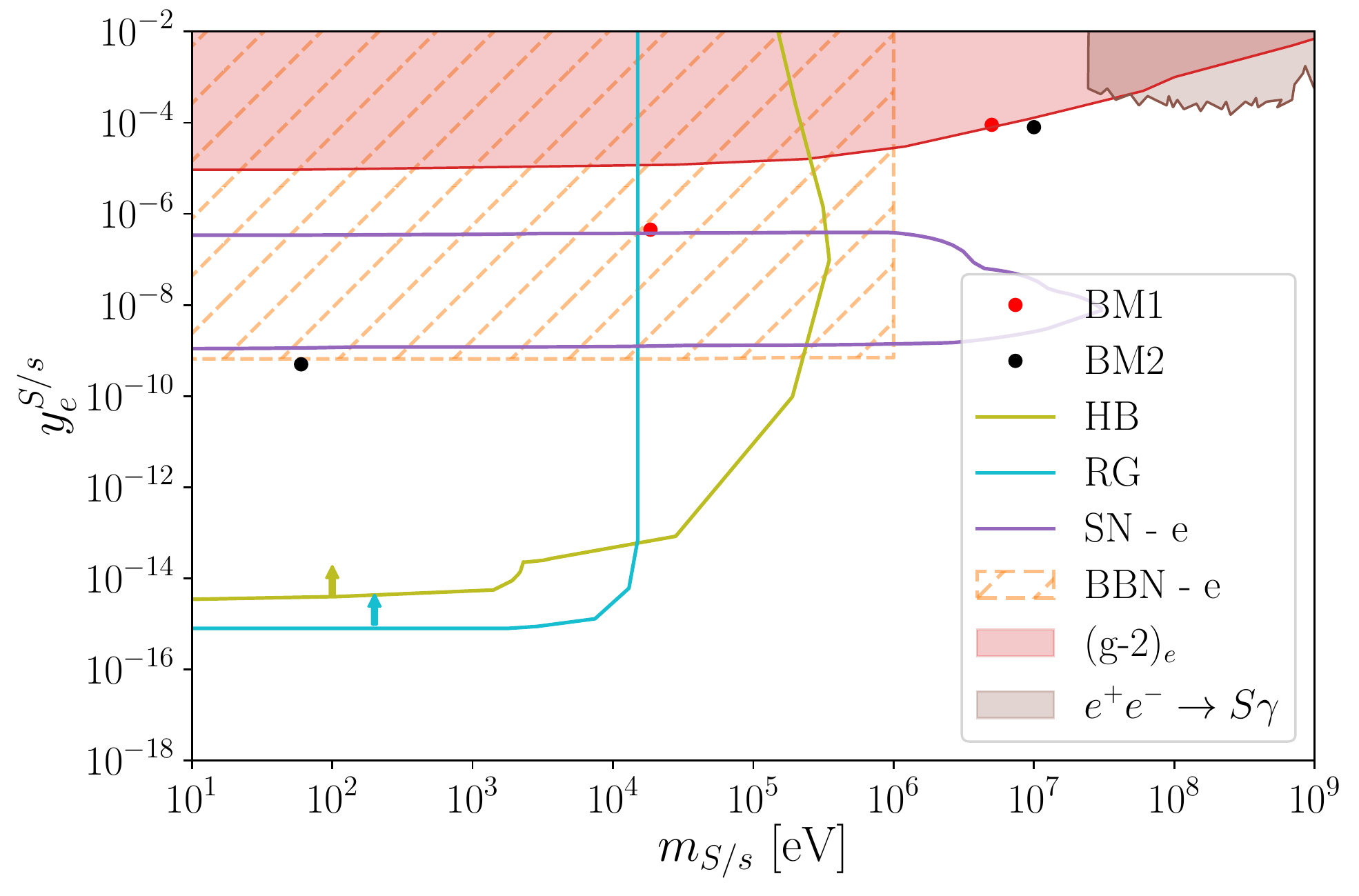}
    \caption{Constraints in the $m_{S/s}-y_e^{S/s}$-plane from \cite{Knapen:2017xzo} and our own analysis, including our two BM points. For a discussion of the various limits and their shading see the main text.}
    \label{fig:exclusions_e}
\end{figure}

\begin{figure}
    \centering
    \includegraphics[width=.48\textwidth]{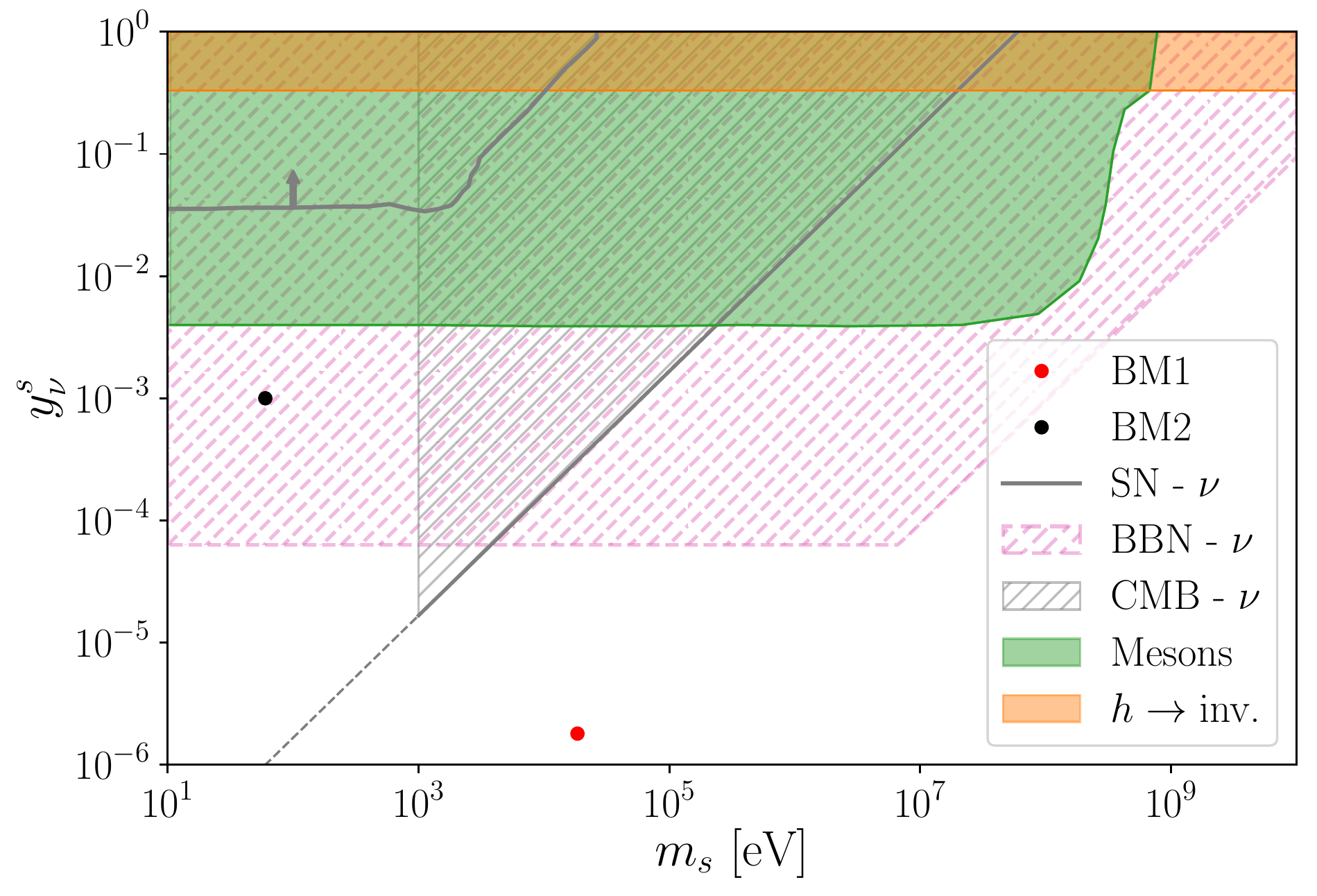}
    \caption{Constraints in the $m_{s}-y_\nu^{s}$-plane including our two BM points. Note that the heavier mediator has too small couplings to appear.}
    \label{fig:exclusions_nu}
\end{figure}

We now confront the neutrino-scattering explanation of the XENON1T excess with various experimental constraints (in our specific setup and in a more general EFT), which are collected in Figs \ref{fig:exclusions_e} and \ref{fig:exclusions_nu} for the scalar-couplings to electrons and neutrinos, respectively.

{\bf Bounds on the combination $y^s_e y^s_\nu$:}
Neutrino-electron scattering has long been a staple signature in experiments aiming to observed solar and reactor neutrinos. These experiments probe very similar physics and place an upper bound on the combination of couplings relevant for the XENON1T signal. Conventionally, bounds on new physics that leads to a recoil spectrum peaking at low energies are interpreted in terms of a neutrino magnetic moment $\mu_\nu$. Currently, the best  limits are from Borexino and GEMMA and stand at $2.9 \times10^{-11} \mu_B$ \cite{Borexino:2017fbd,Beda:2013mta}. This is right on the edge of the values preferred by the XENON1T excess, $\mu_\nu = 1.4 \-- 2.9\times 10^{-11} \mu_B$ but does not exclude the magnetic moment interpretation \cite{Aprile:2020tmw}. This observation is highly relevant for the light scalar mediation scenario under consideration here. In the  energy range where the XENON1T signal is observed the recoil energy distribution of events that are induced by solar or reactor neutrinos interacting via a light scalar ($m_s \lesssim E_r$) or a magnetic moment are essentially indistinguishable. Consequently, an interpretation of the Borexino data in our model will lead to a constraint that is just on the upper boundary of the preferred region. With the signal and the expected exclusion so close to each other the exact position of the bound will depend on the details of the experimental data and the statistical procedure. A naive phenomenologists recast is therefore unlikely to allow a clear comparison. Thus, we refrain from quoting an explicit limit derived from a reinterpretation of Borexino and GEMMA data and just note that the bound is expected to be closely aligned with the upper edge of the preferred values of $\sqrt{y_e
^sy_\nu^s}$.  

{\bf Bounds on electron couplings:} Beyond the SM forces coupling to the electron can be tested very rigorously with terrestrial  precision experiments. In the mass range of interest here the most stringent constraints come from the anomalous magnetic moment of the electron $a_e$ since both the experimental measurement and the SM prediction are incredibly precise. At $3\sigma$ the deviation of $a_e$ from the SM expectation is limited to $\delta a_e \lesssim 1.4 \times 10^{-12}$ \cite{Hanneke_2008,Giudice:2012ms}. A new scalar contributes \cite{Jackiw:1972jz} 
\begin{equation}
    \delta a^s_e = \frac{(y_e^s)^2}{4\pi^2}\, \frac{m_e^2}{m_s^2}\ I_S\left(\frac{m_e^2}{m_s^2}\right)\,,
\end{equation}
where the loop function is given by 
\begin{equation}
    I_S(r)=\int_0^1\! dz\, \frac{z^2(2-z)}{1-z+z^2 r}\,.
\end{equation}
For $m_s\ll m_e$ this leads to $y^s_e \lesssim 10^{-5}$ 
while the limit relaxes for $m_s\geq m_e$, c.f.~Fig.~\ref{fig:exclusions_e}. Softer terrestrial constraints can be derived from $e^+e^-$ colliders through the process $e^+e^- \to \gamma s$. They have the largest impact close to $m_s \sim {\cal O}(1) $GeV, see for example \cite{Knapen:2017xzo}. 

In addition, there are a number of bounds on $y_e^{s,S}$ from astrophysical and cosmological observations. 
If the mass of the mediator is comparable or smaller than the core temperature of a star, the emission of the scalars can contribute to the energy loss and change the properties and dynamics of these astrophysical systems. Strong limits can be derived from  red giants (RG) and horizontal branch stars (HB). We adopt the results of \cite{Hardy:2016kme,Knapen:2017xzo}  where plasma mixing is considered to be the main production mechanism of the light scalars, for a more recent analysis of the impact of stellar cooling on new physics in other models see \cite{Capozzi:2020cbu}. 
In principle, for $m_{s,S}\ll 10$ keV the RG bound excludes couplings larger than $y_e^s \approx 10^{-15}$. Clearly, such a small $y_e^s$ would prevent a solar neutrino interpretation of the XENON1T excess for all reasonable values of $y_\nu^s$. 
The bounds from observations of HB stars are less sever at low masses but take over for $m_{s,S}\gtrsim 10$\,keV.
However, it is conceivable that these constraints can be circumvented in the presence of additional new physics such as an environment-dependent mass for the scalar \cite{DeRocco:2020xdt} similar to the chameleon mechanism considered in cosmology \cite{Khoury:2003aq,Joyce:2014kja}. 
Alternatively, the coupling of the light scalar $s$ to electrons could depend on the matter background and may be suppressed in a high density environment such as stellar cores. We will briefly sketch a concrete realization  of this possibility in Sec.~\ref{Sec:evade_astro_bounds}.
Therefore, we consider such astrophysical bounds less robust than the direct laboratory bounds discussed before and in consequence draw them as lines, removing the shading from the disfavored regions.

Another constraint for mediator masses up to ${\cal O}(10)$~MeV is set by the supernova (SN) SN1987A, as additional light degrees of freedom would rapidly cool the SN in contrast to observation \cite{Raffelt:1996wa}. Due to the very high density of the SN core the scalar mediator can be trapped before actually leaving the core. 
We consider the limits from \cite{Knapen:2017xzo}, where only the resonant production via mixing with the longitudinal component of the photon is included and direct production through Compton scattering or electron-ion recoil is neglected.
This is possible for $m_s < w_p \sim 20$~MeV, where $w_p$ is the photon plasma frequency \cite{Chen_2018}. The trapping regime for resonant production is also included by using the balance of production and absorption rate, with the requirement of the scalar to be re-absorbed in a range of $R\approx10$~km. In this trapping regime, the decay $s \to e^+e^-$ determines the bound for masses MeV$\leq m_s \leq 30$ MeV. 

In addition, there are bounds from Big Bang Nucleosynthesis (BBN) for additional light degrees of freedom entering thermal equilibrium with $e$ and $\gamma$. On top of an increase of $N_{eff}$, the entropy release from $e^+e^-$ annihilation is diluted in that case. This leads to a lower photon temperature during BBN and therefore an higher baryon-to-photon ratio, which causes a decrease of the deuterium abundance \cite{Knapen:2017xzo}. 
For $m_s \lesssim 1$ MeV The  BBB bound is largely flat and requires $y^s_e\lesssim 10^{-9}$.
Even though the BBN bound is quite robust it can be circumvented in our setup. A late time phase transitions in the new physics sector can prevent the mixing of $s$ and $S$ in the early Universe and thus remove the coupling between the lighter scalar and the electrons at the relevant temperatures. We will comment more on this possibility in the next section.

{\bf Bounds on neutrino couplings:}
New physics interacting with neutrinos is harder to test than in the case of electrons and we expect the bounds to be less constraining. 
Robust terrestrial constraints arise from searches for new meson decays such as $K^-/D^-/\pi^-\to e^- s \nu$~\cite{Berryman:2018ogk}.  Alternatively, also decays to $\mu^-$ can be considered.  We show the strongest combination of those in Fig.~\ref{fig:exclusions_nu} assuming a flavor universal coupling. In case of flavor non-universality, the bounds for electron couplings are slightly stronger.

For $m_h> m_s \gtrsim 1$\,GeV limits on the decay width of Higgs to invisible states via $h\to s \nu\nu$ give the strongest bound on $y_\nu^s$ \cite{Berryman:2018ogk}. In Fig.~\ref{fig:exclusions_e} we use the latest ATLAS result of BR$(h\to \text{inv.}) < 0.13$ \cite{ATLAS:2020cjb}.

The observation of MeV-scale neutrinos originating from SN1987A constraints the neutrino self-interaction~\cite{ Shalgar:2019rqe}. Scattering of the SN-neutrinos with the C$\nu$B via the new mediator shifts their energy to significantly lower values and potentially brings them below the detection threshold. In addition the SN neutrinos get deflected which delays their arrival on earth. An early bound was derived in \cite{Kolb:1987qy}, we show the one from \cite{Shalgar:2019rqe} in Fig.~\ref{fig:exclusions_nu}, where the recent limits on the neutrino masses were used.

The model could also have an impact on the amount of radiation in the Universe which can be tested by BBN. In particular the right-handed neutrinos are dangerous since fully thermalized each of them will contribute $\Delta N_{eff}=1$ while the  upper bound stands at $\approx 0.2$~\cite{Cyburt:2015mya}. Therefore, only the region of parameter space were the right-handed neutrinos do not reach thermal equilibrium before the left-handed ones decouple from the SM bath are allowed by cosmology. Even if the initial population of $\nu_R$ is negligible we can produce them in neutrino-antineutrino scattering via t-channel $s$ exchange. A good estimate for thermalization can be obtained by requiring that the production rate $\gamma$ does exceed the Hubble rate $H$ prior to neutrino decoupling which happens at about $2\--3$\,MeV. In our model the thermally averaged production rate reads
\begin{align}
    \gamma \approx\langle \sigma v \rangle \times n_{\nu} \approx \frac{(y^s_\nu)^4}{512\pi}\,T,
\end{align}
where $n_{\nu}$ is the equilibrium number density of neutrinos and $\langle \sigma v \rangle$ is the thermally averaged $\nu_R$ production cross section. By equating the rate and $H$ we find $y^s_{\nu}\lesssim 6.3 \times 10^{-5}$ for $m_{s}\ll 2$~MeV. For larger masses the bound weakens. The contribution of $s$ is less pronounced than in the case of electrons since the absence of a $\nu_R$ bath prevents the direct production of $s$. We note that this bound can be avoided if an additional mass terms for the right-handed neutrinos makes them too heavy to contribute to $N_{eff}$. This can be realized rather straightforwardly in our setup by increasing $v_\nu$ such as to generate a more sizable Dirac-mass term that then leads to viable neutrino masses via see-saw suppression in the presence of large Majorana masses for the right-handed neutrinos. This would provide a hybrid explanation for the smallness of neutrino masses, which will however require a refined analysis, that goes beyond the scope of this work.
 
Finally, there are constraints from CMB. If the interaction rate of neutrinos is high enough they cannot be treated as a free-streaming gas and the impact of their interactions has be to included in the Boltzmann equations governing the evolution of the primordial perturbations. For a heavy mediator this leads to an upper bound on the interaction strength of $\left(y^s_\nu/m_s\right)^2\leq (0.06 \,\mbox{GeV})^{-2}$ \cite{Archidiacono:2013dua}. In order for this estimate to be valid we need $m_s\gg 10$ eV and, therefore, the limit becomes unreliable towards the lower end of the mass range considered here. \footnote{Alternative limits on very light mediators are also available~\cite{Archidiacono:2013dua} but they only become applicable at even smaller masses.} 

{\bf Benchmark models:} In order to confront our model for the XENON1T excess with these astrophysical and laboratory constraints, let us define two different benchmarks (BMs) that both deliver a good fit to the anomaly as in Eq.~\eqref{eq:fit}. While we require roughly natural scales for the model, we are mainly led by the goal to avoid the most severe experimental bounds. The BMs are defined by the independent input parameters\vspace{2mm}

\begin{tabular}{c|c|c|c|c|c|c}
    BM & $M_\nu$ & $M_e$ & ${\rm y}_\nu^{\cal S}$ &  ${\rm y}_e^{\cal S}$ & $\Lambda$ & $\lambda_{\nu e}$ \\\hline
    BM1 & 18.5\,keV & 5\,MeV & $1\!\times\!10^{-4}$ & 0.005 & 10\,TeV & $3\!\times\!10^{-4}$ \\\hline
    BM2 & 60\,eV& 10\,MeV & 0.06 & 0.005 & 10\,TeV & 0.001
\end{tabular} \vspace{2mm}

\noindent which lead to the vevs  $(v_\nu,v_e)=(26.5\,{\rm keV},5.3\,{\rm GeV})$  and $(v_\nu,v_e)=(50\,{\rm eV},5.9\,{\rm GeV})$ for BM1 and BM2, respectively, as well as to the (derived) physical couplings \vspace{2mm}

\begin{tabular}{c|c|c|c|c}
BM & $y^s_\nu$ & $y^s_e$ & $y^S_\nu$ & $y^S_e $\\\hline
BM1 & $1.8\!\times\!10^{-6}$  & $-4.5\!\times\!10^{-7}$ & $8.3\!\times\!10^{-9}$ & $9\!\times\!10^{-5}$\\\hline
BM2 & 0.001 & $-5\!\times\!10^{-10} $ & $6\!\times\!10^{-9}$ & $8\!\times\!10^{-5}$
\end{tabular}\vspace{2mm}

\noindent with mixing angles $s_\theta^1= -5\times10^{-3}$ and $s_\theta^2=-6\times10^{-6}$. We note that in defining these benchmarks we followed two different assumptions regarding the neutrino masses, that are both consistent with the values above.\ \,
1)~We assumed an `inverted' neutrino-mass hierarchy with $m_{\nu^3} \ll m_{\nu^1} \sim m_{\nu^2} \sim 0.05\,$eV. In this case actually both $\nu^{1,2}$ couple to the mediator $s$ with similar strength $y^s_\nu$, while the interaction of the lightest neutrino is negligible, see Eq.~\eqref{eq:YukFin2}. Since $\nu^{1,2}$ contain almost all the electron-flavor content and couple universally to $s$,
basically no flux from the sun will be lost when considering neutrino-electron scattering in XENON1T and the analysis as described above remains valid.\ \, 2)~A `normal' hierarchy with  $m_{\nu^1} \sim m_{\nu^2} \sim 0.05\,{\rm eV} \ll  m_{\nu^3}$,  would also be consistent with the same BMs, where we now assume that both chiralities of the heaviest state $\nu^3$ are even under the $Z_2^{\nu}$ symmetry, such that it does not couple to $s$ (while again the electron-neutrino content is almost entirely in the universally coupling eigenstates).

For both BMs, we arrive at a prediction for the strength of the anomaly of
\begin{equation}
\label{eq:res}
    y^s_e\, y^s_\nu \approx  - (5-7) \times 10^{-13}\,,
\end{equation}
in line with the best-fit value obtained before in Eq.~\eqref{eq:fit}.\footnote{Moreover, both BMs satisfy the positive-definiteness condition $M_\nu M_e > 2 \lambda_{\nu e} v_\nu v_e$, ensuring a proper potential minimum.}
Finally, the couplings associated with BM1 (BM2) are displayed in red (black) in the landscape of collected bounds on $y_e^{s/S}$ and $y_\nu^s$ in Fig.~\ref{fig:exclusions_e} and Fig.~\ref{fig:exclusions_nu}, respectively.

A few comments are in order. 
First, a value of $v_e > M_e$, somewhat above the electron mass, leads to a coupling of electrons to the heavy mediator $S$ ($y_e^S\sim 10^{-4}$) that just evades the precision bounds for the corresponding mediator mass of $m_S \sim 10$\, MeV \cite{Knapen:2017xzo}.
On the other hand, the coupling to the potentially dangerously light $s$ is suppressed in $s_\theta$, pushing the resulting interaction {\it just} into the window above the SN1987a exclusion region but below the $(g-2)_e$ limit for BM1, while BM2 can even evade BBN constraints without further ingredients (at the price of a higher neutrino coupling). 
The BBN constraint for electrons in BM1 can be avoided via a late phase transition, generating the vev $v_\nu>0$ below $T\approx 150$\,keV, as we discuss in the next section.

\begin{figure}    \centering
    \includegraphics[width=.48\textwidth]{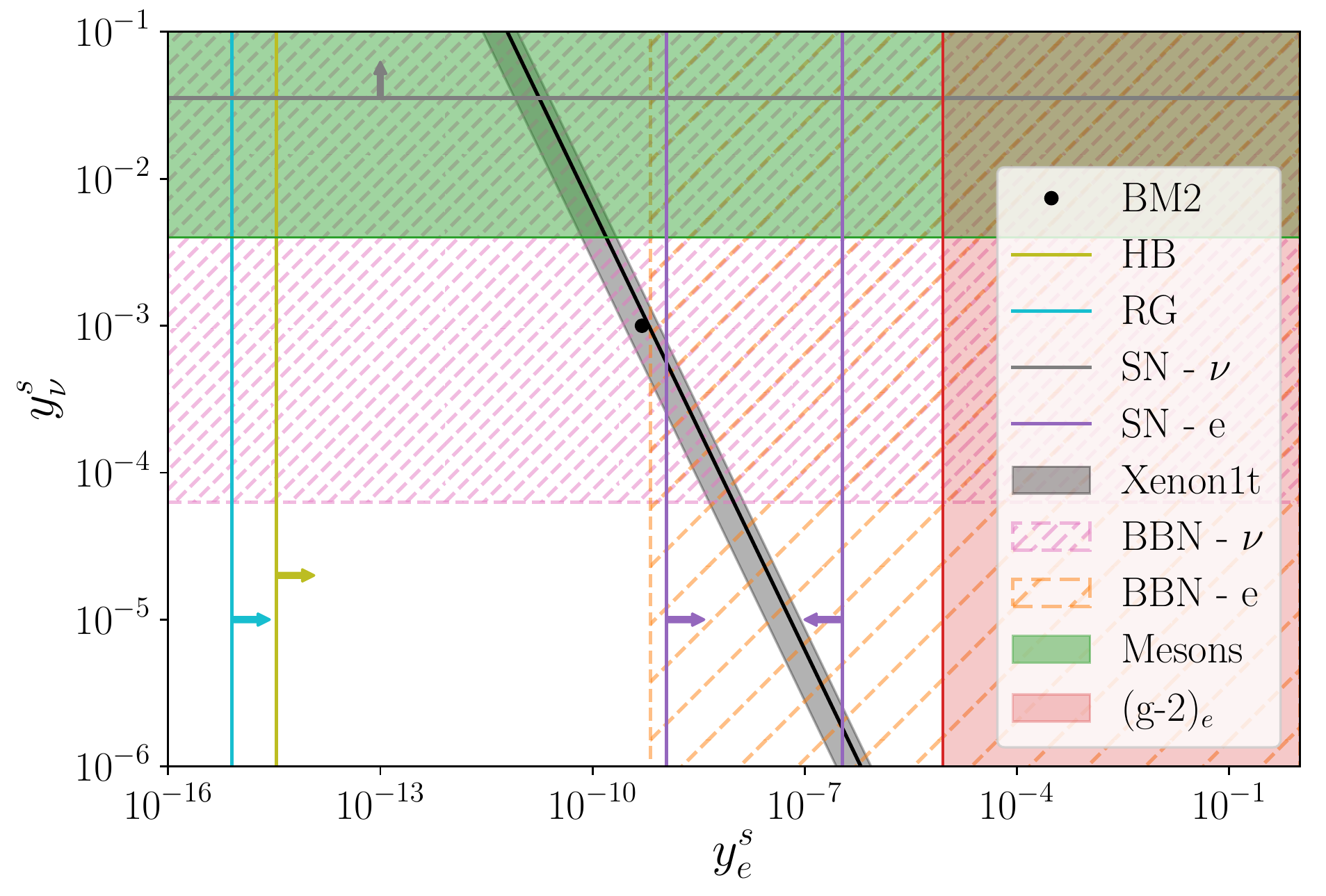}
    \caption{Constraints in the $y_e^s-y_\nu^s$-plane for a $60$~eV mediator (as in BM2) and the $1\sigma$ preferred region from our fit to the XENON1T excess. The BBN bound on the electron coupling, indicated by the hatched region, can be circumvent by a late time phase transition.}
    \label{fig:exclusions_e_nu_60eV}
\end{figure}

\begin{figure}    \centering
    \includegraphics[width=.48\textwidth]{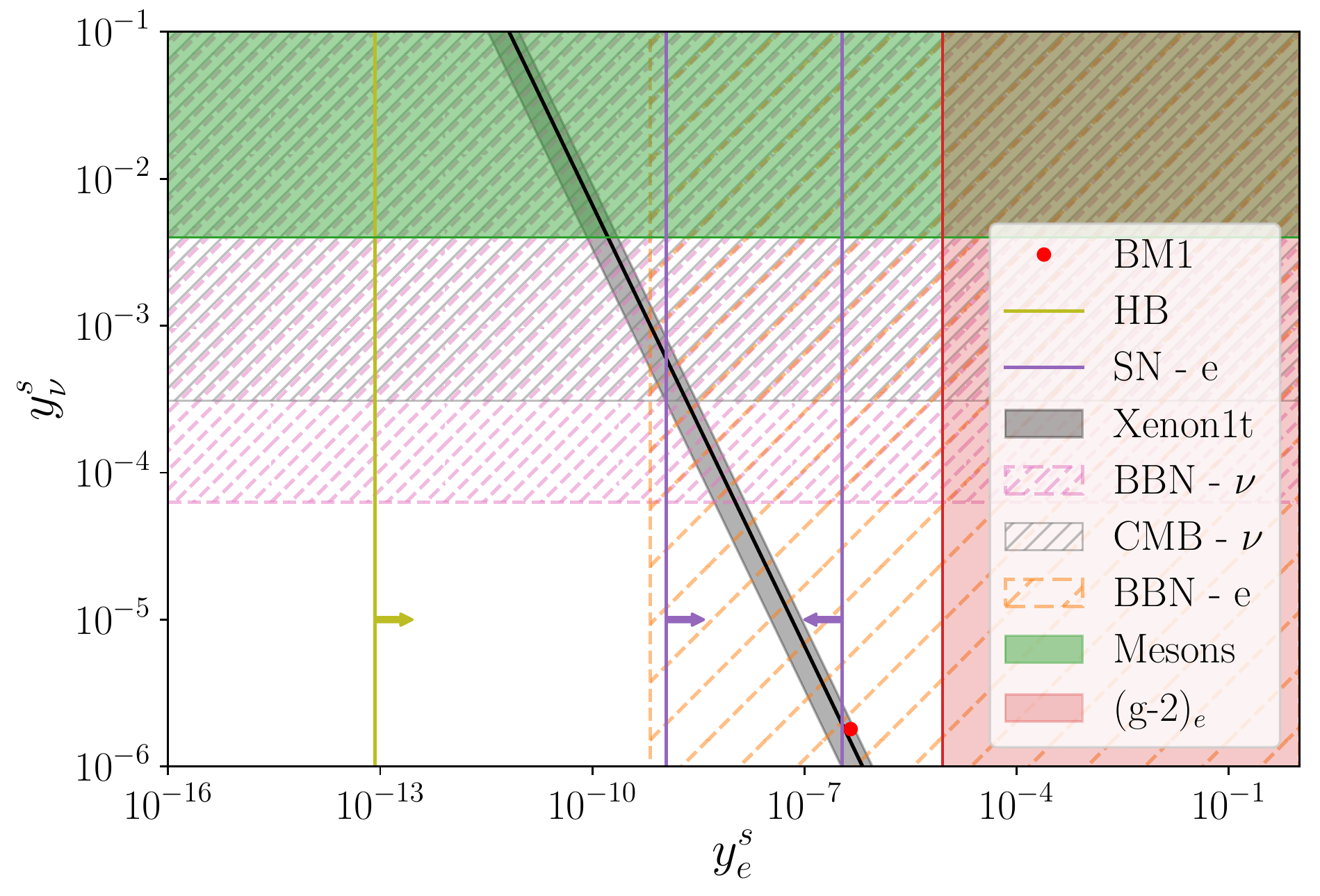}
    \caption{Constraints in the $y_e^s-y_\nu^s$-plane for a $20$~keV mediator (as in BM1) and the $1\sigma$ preferred region from our fit to the XENON1T excess. The BBN bound on the electron coupling, indicated by the hatched region, can be circumvent by a late time phase transition.}
    \label{fig:exclusions_e_nu_20keV}
\end{figure}

{\bf Free EFT description:}
Finally, we confront the general EFT resolution to the XENONT1T anomaly via scalar couplings to electrons and neutrinos in the couplings plane with the constraints discussed above. Being agnostic, here we just employ the effective Lagrangian (omitting kinetic and potential terms)
\begin{equation}
\label{eq:EFT2}
    \mathcal{L}_{\rm eff} = -\frac{\sqrt 2}{v} \Big[y_\nu^s\, \bar{L}_L^1 H \nu^1_R\, s + y_e^s \bar{L}_L^1 H e_R\,s + \mathrm{h.c.}\Big]\,,
\end{equation}
which can be obtained from Eq.~\eqref{eq:LEFT} by neglecting the second scalar singlet, while coupling the remaining one to both electrons and neutrinos and removing the $Z_2$ symmetries as well as the vev of the mediator. In consequence, all fermion masses are solely induced by the Higgs and $y_{\nu,e}^s$ are now completely free couplings. In particular, Eq.~\eqref{eq:EFT2} corresponds to a subset of operators of the general \eDMEFT\ \cite{Alanne:2017oqj,Alanne:2020xcb}.

In Fig.~\ref{fig:exclusions_e_nu_60eV} we show the constraints and best fit region in the $y_e^s-y_\nu^s$-plane for a mediator mass of $60$~eV and in Fig.~\ref{fig:exclusions_e_nu_20keV} for $20$~keV respectively. For comparison we also add the coupling values used in the two BMs above.

There are two regions in the couplings preferred by the XENON1T fit, that potentially remain valid but both need extra mechanisms to avoid bounds from BBN in the early universe. The one around $y_e^s =\mathcal{O}(10^{-9})$ is excluded by the neutrino BBN bound. As discussed before, this could be avoided by an additional mass terms for the right-handed neutrinos. The other benchmark around $y_e^s =\mathcal{O}(10^{-6})$ is under pressure from the electron BBN bound. However, here a late phase transition can remove the interaction of the light mediator and electrons during the relevant age of the Universe and make this point potentially viable.

\section{Evading astrophysical bounds}
\label{Sec:evade_astro_bounds}
In this section we will introduce a mechanism to alleviate the astrophysical bounds, based on a new scalar singlet $\phi$ with a potential similar to the one employed by \cite{DeRocco:2020xdt}. In contrast to this work, we aim to remove the vev of $S_\nu$ and not to increase the mass of the offending field.  The new scalar couples to the baryons with a $g \phi N \bar N$ operator and has a four scalar interaction with the light, neutrinophilic scalar $S_\nu$. The relevant part of the Lagrangian reads
\begin{align}
\mathcal{L} &\supset \frac{1}{2} \left( \partial_\mu \phi \partial^\mu \phi  -  m^2_\phi \phi^2 \right) + g \phi N \bar{N} - \frac{1}{4} \lambda_\phi \phi^4 \\
+&  \frac{1}{2} \left( \partial_\mu S_\nu \partial^\mu S_\nu  -  \tilde \mu_\nu^2 S_\nu^2 \right) - \frac{1}{4} \lambda_\nu S_\nu^4 - \lambda_{\phi S} \phi^2 S_\nu^2 \, \nonumber
\end{align}
where $g$ is the coupling between $\phi$ and baryons, here denoted by $N$, $\lambda_{\phi S}$ is a four scalar interaction, and we have defined the effective $\tilde \mu_\nu^2 \equiv \mu_\nu^2 + 2 \lambda_{\nu e} v_e^2$.
The term $g \phi N \bar{N}$  induces a background dependent contribution to the $\phi$ potential. In a bath $N \bar{N} \rightarrow \frac{n_B}{\gamma}$, where $\gamma\approx1$ in the non-relativistic backgrounds of interest to us. 
Taking $m_\phi^2$ to be positive and neglecting $\lambda_{\phi S}$ for the moment the potential develops a vacuum expectation value $v_\phi = g\, n_B /m_\phi^2$ provided that $\lambda_\phi$ is small enough. For non-vanishing $\lambda_{\phi S}$  the  vacuum expectation value influences the $S_\nu$ sector and shifts the effective $\tilde{\mu}_{\nu}^2$ term to $\tilde{\mu}^2= \tilde \mu_\nu^2 + 2 \lambda_{\phi S} (g n_B/ m^2_\phi)^2$. If the new physics contribution exceeds $|\tilde \mu_\nu^2|$ the effective parameter turns positive and the vev of $S_\nu$ disappears. This removes the mixing with the electrophilic scalar $S_e$ and shuts off the coupling between $s$ and the electrons that drive the astrophysical constraints. 

The mechanism looks promising but one might wonder if parameters that achieve the desired results can be found without violating experimental bounds. Therefore, we will provide one working assignment of the couplings and masses that allows to evade the astrophysical bounds for BM2. HB stars possess a central density of about
 $10^4 \, \mbox{g}/\mbox{cm}^3$~\cite{Hardy:2016kme}\footnote{RG stars are even denser and our argument goes through without modification.} which induces
 $v_\phi \approx g\, 4\times 10^{-5} \mbox{MeV}^3 /m_\phi^2$. Choosing
 $m_\phi=1 \,\mbox{eV}$ and $g= 10^{-11}$ leads to
 $v_\phi \approx 400$~eV. The values for $m_\phi$ and $g$  respect bounds from searches for a fifth force \cite{Banks:2020gpu}. 
 In our benchmark scenario $|\tilde \mu_\nu| \approx 40$\,eV which implies that $\lambda_{\phi S}\approx 0.01$ is sufficient to remove the minimum 
 at non-zero $v_\nu$. In principle, one should consider the whole potential simultaneously and not treat
  the potentials consecutively. However, in the case at hand, $v_\phi$ exceeds $v_\nu$ significantly and thus the expected influence of 
  $S_\nu$ on $\phi$ is small for the values of the coupling considered here. Nevertheless, we have checked explicitly that our argument goes through with minimal corrections if the full potential is considered.    

\section{Avoiding BBN Bounds via a Late Phase Transition}
\label{sec:PT}
As discussed before, without further ado, BM1 would be excluded from BBN bounds on the electron coupling. However, in this section we will demonstrate how our scenario naturally realizes a late $Z_2$ breaking phase transition delaying the coupling of the electron to the light mediator until after BBN has completed. The scalar potential of our model can lead to a rich cosmological history in which the $Z_2^{\ell}$ symmetries are broken in a stepwise fashion \cite{Carena:2019une}. For simplicity we neglect mixing between the scalars $S_\nu$, $S_e$ and the Higgs doublet $H$ by turning $\lambda_{HS_\ell}$ zero. The tree-level scalar potential is then given by
\begin{equation}
    V_{\textrm{tree}}=\frac{1}{2}\mu_\nu^2S_\nu^2+\frac{1}{2}\mu_e^2S_e^2+\lambda_{\nu e}S_\nu^2S_e^2+\frac{1}{4}\lambda_\nu S_\nu^4+\frac{1}{4}\lambda_eS_e^4\,.
\end{equation}

To study the cosmological evolution of this potential we add the one-loop thermal corrections given by \cite{Quiros:1999jp}
\begin{equation}
    V_{\textrm{thermal}}= \frac{T^4}{2\pi^2} \left(J_B\!\left(\frac{m_s^2}{T^2}\right)+ J_B\!\left(\frac{m_S^2}{T^2}\right)\right)\,,
\end{equation}
where $J_B(\alpha)=\int_0^{\infty}x^2\ln(1-e^{\sqrt{x^2+\alpha}})dx$ is the thermal correction for bosonic degrees of freedom. Working in the high-temperature limit, the thermal corrections have analytical forms  $J_B(\alpha)=-\frac{\pi^4}{45}+\frac{\pi^2}{12}\alpha+O(\alpha^{3/2})$. Note that, since mixing between $S_\nu$ and $S_e$ is small, we can take approximately $m_s\approx M_\nu$ and $m_S\approx M_e$. Under these approximations, the critical temperature $T_{c2}$ at which a second minimum $(\langle S_\nu\rangle, \langle S_e\rangle)=(0,v_e)$  degenerate with the $Z_2^\nu \times Z_2^e$ preserving vacuum $(\langle S_\nu\rangle, \langle S_e\rangle)=(0,0)$ forms, is given by
\begin{equation}
    T_{c2}=\frac{\sqrt{-12\mu_e^2}}{2\lambda_{\nu e}+3\lambda_e}\,.
\end{equation}
A second phase transition appears once the temperature has dropped to $T_{c1}$ at which a non zero vev of $S_\nu$ forms, with
\begin{equation}
    T_{c1}^2=\frac{12\big(2\lambda_{\nu e}\mu_e^2-\lambda_e\mu_\nu^2\big)}{\lambda_e(2\lambda_{\nu e}+3\lambda_\nu)-2\lambda_{\nu e}(2\lambda_{\nu e}+3\lambda_e)}\,.
\end{equation}
For BM1 the first phase transition occurs around $500$ MeV while the second phase transition occurs at $150$ keV. At this temperature most of the photon heating is complete and the electron density has already dropped significantly -- and, therefore, the thermalization rates are starting to be exponentially suppressed. 

\section{Conclusions}
\label{sec:Conc}
We have investigated the excess in low energy electron recoil events reported by the XENON1T collaboration. Our work is based on the $Z_2$ symmetric extended DM EFT which connects neutrino mass generation and DM \cite{Goertz:2019wtt}. We find that conventional DM-electron scattering only allows for a marginally better fit than the background-only hypothesis since the signal spectrum peaks at lower energies than observed experimentally. Therefore, DM does not provide convincing explanation of the data. However, the new neutrino and electron couplings induced by the neutrino mass mechanism embedded in the model predict a significant neutrino-electron scattering cross section. Including this interaction in the fit improves it considerably and we find that a light scalar with an average electron-neutrino coupling of $\sqrt{y^s_e y^s_\nu}\approx 7.9 \times 10^{-7}$ is preferred by more than~2\,$\sigma$.  

These observations motivated us to scrutinize the parameter space  of the model in more detail and compare it to limits from various other observations. In general the parameter space that allows for a successful explanation of the XENON1T excess is rather constrained. While limits from terrestrial experiments can be avoided comparatively easily, bounds from cosmology are more constraining. In particular BBN bounds on a light scalar coupling to electrons are very severe.   Interestingly, the model under consideration here naturally allows for a late phase-transition in the early Universe which prevents the scalar-electron coupling during BBN. However, in such a scenario additional contribution to the right-handed neutrino masses are required in order to avoid their thermalization prior to BBN. Once this is taken into account we find solutions that comply with cosmological bounds. There remains a strong tension with astrophysics bounds, that rely on stellar cooling arguments. Nevertheless, if we allow for the possibility of additional new physics, these limits can be avoided by a density dependent contribution to the $S_\nu$ potential that suppresses or even removes the mixing with $S_e$ in a high density environment. 
 
All considered, a new physics explanation of the excess is a tantalizing possibility, but in light of stringent constraints from other observations this potential sign of physics beyond the Standard Model should be taken with a grain of salt. Luckily, the upcoming run of the XENONnT experiment will be able to weigh in on this question in the near future and either strengthen the excess or rule it out conclusively.

\section{Acknowledgments}
\label{sec:Ack}
We thank Tommi Alanne and Simone Blasi for helpful discussions.
VT acknowledges support by the IMPRS-PTFS and KMTN acknowledges support from  the research training group "Particle Physics Beyond the Standard Model" (Graduiertenkolleg 1940).

\appendix
\section{Potential UV completions}
\label{sec:UV}

In this appendix, we present two explicit UV completions that lead to the relevant terms in the Lagrangian~\eqref{eq:LEFT} after integrating out additional heavy degrees of freedom.

\subsubsection*{Model 1: 3HDM + Scalar Singlets}
Consider a model with 3 electroweak doublets, the SM-like Higgs $H$ as well as two additional
doublets $H_e$ and $H_\nu$, with $Z_2^e=-1$ and $Z_2^\nu=-1$, respectively.
A minimal Lagrangian, respecting the $Z_2$ symmetries, reads (omitting flavor indices)
\begin{equation}
\begin{split}
    {\cal L} \supset & - y_e^H \bar L_L H_e e_R - y_\nu^H \bar L_L \tilde H_\nu \nu_R \\
    & -\lambda_e m_{H_e} H^\dagger H_e S_e -\lambda_\nu m_{H_\nu} H^\dagger H_\nu S_\nu +  {\rm h.c.} \\
    & -m_{H_e}^2 |H_e|^2 - m_{H_\nu}^2 |H_\nu|^2 \,.
\end{split}
\end{equation}
After integrating out $H_e$ and $H_\nu$ at tree level, the required Yukawa couplings, sought to explain the XENON1T anomaly, are generated with $y_e^{\cal S} \sim y_e^H \lambda_e$ and $ y_\nu^{\cal S} \sim y_\nu^H \lambda_\nu$. Since we consider $\Lambda \sim m_{H_{e,\nu}} \sim 10\,$TeV, together with small dimensionless couplings, the scenario is expected to be safe from further experimental constraints.

\subsubsection*{Model 2: Vector-Like Leptons + Scalar Singlets}
As a second option, we add a set of vector-like leptons $E$ and $N$, with the quantum numbers of the right-handed electron
and neutrino, respectively (but no $Z_2$ charge), to the setup.
The corresponding Lagrangian becomes
\begin{equation}
\begin{split}
{\cal L} \supset& - y_E \bar L_L H E_R - y_N \bar L_L \tilde H N_R  \\ &- y_E^S \bar E_L S_e e_R - y_N^S \bar N_L S_\nu \nu_R  
+  {\rm h.c.} \\ & - m_{E} \bar E E - m_{N} \bar N N \,.
\end{split}
\end{equation}
Again, after integrating out $E$ and $N$, the sought Yukawa couplings are obtained straightforwardly with $ y_e^{\cal S} \sim y_E y_E^S$, $ y_\nu^{\cal S} \sim y_N y_N^S$, and $\Lambda \sim m_{E,N}$.
Similarly to Model~1, we expect the setup to be phenomenologically viable given the large suppression scale $\Lambda$.

\bibliographystyle{apsrev4-1}
\bibliography{xenon1Tee}

\end{document}